\documentclass[preprint,12pt]{elsarticle}
 \usepackage{amsmath,amssymb}
 \usepackage{epsfig,graphicx}
 
\journal{Physical Letters B}
 
\begin{document}
 
\begin{frontmatter}

\title{\bf Extraction of Kaon Formfactors from $K^-\to\mu^-\bar\nu_\mu\gamma$ Decay at ISTRA+ Setup}
\author{V.A.~Duk \fnref{fn1} }
\author{V.N.~Bolotov,  V.A.~Lebedev, A.A.~Khudyakov,\\  A.I.~Makarov, A.Yu.~Polyarush,  V.P.~Novikov}

\address{Institute for Nuclear Research of RAS, Moscow, Russia}

\author{S.A.~Akimenko, G.I.~Britvich, A.P.~Filin, A.V.~Inyakin,\\
I.Ya.~Korolkov, V.M.~Leontiev, V.F.~Obraztsov, V.A.~Polyakov,\\
V.I.~Romanovsky, O.G.~Tchikilev, V.A.~Uvarov, O.P.~Yushchenko}
\address{Institute for High Energy Physics, Protvino, Russia}

\fntext[fn1]{Viacheslav.Duk@cern.ch}


\begin{abstract}
The radiative decay $K^-\to
\mu^-\bar\nu_{\mu} \gamma$   has been studied at ISTRA+ setup in a 
new kinematic region. About 22K
events  of $K^-\to\mu^-\bar\nu_{\mu} \gamma$ have been observed. 
The sign and value of $F_{V}-F_{A}$ have been measured for the first time. The  result
is $F_{V}-F_{A}=0.21\pm0.04(stat)\pm0.04(syst)$.

\end{abstract}


\end{frontmatter}

\section{Introduction}

Radiative kaon decays are dominated by long distance (low energy) physics. For low energy
 processes there are no direct predictions from SM and effective theories such
 as Chiral perturbation theory ($\chi$PT) are used. $\chi$PT gives quantitative predictions for most kaon decay
 modes. That is why radiative kaon decays provide a testing ground for $\chi$PT.
Moreover, these decays are sensitive to New Physics. 

 The decay  $K^-\to\mu^-\bar\nu_{\mu} \gamma$ is sensitive to hadronic weak
currents in low-energy region.
 The decay amplitude includes two terms:  internal bremsstrahlung
(IB) and structure dependent term (SD). IB contains radiative corrections from
$K^-\to\mu^-\bar\nu_{\mu}$. SD allows to probe electroweak 
structure of kaon.

The differential decay rate can be written
in terms of standard kinematic variables $x=2 E^\star_\gamma/M_k$
and $y=2E^\star_\mu /M_k$ (see \cite{rate} for details), $E^\star_\gamma$ being photon energy
 and $E^\star_\mu$ muon energy in cms. It includes IB, SD$^\pm$ parts
and their interference INT$^\pm$. The SD$^\pm$ and INT$^\pm$ contributions are determined by
two formfactors $F_{V}$ and $F_{A}$.  

The general formula for decay rate is as follows:\\

$\frac{\mbox{$d\Gamma$}}{\mbox{$dxdy$} }  = A_{IB}f_{IB}(x,y) + A_{SD}[(F_V+F_A)^2f_{SD^+}(x,y) + (F_V-F_A)^2f_{SD^-}(x,y)] \\
-A_{INT}[(F_V+F_A)f_{INT^+}(x,y) + (F_V-F_A)f_{INT^-}(x,y)] $\\

where \\

$f_{IB}(x,y)=[\frac{\mbox{$1-y+r$}}{\mbox{$x^2(x+y-1-r)$}}][x^2 + 2(1-x)(1-r) - \frac{\mbox{$2xr(1-r)$}}{\mbox{$x+y-1-r$}}],$ \\ 

$f_{SD^+}(x,y)=[x+y-1-r][(x+y-1)(1-x)-r],$ \\

$f_{SD^-}(x,y)=[1-y+r][(1-x)(1-y)+r],$ \\

$f_{INT^+}(x,y)=[\frac{\mbox{$1-y+r$}}{\mbox{$x(x+y-1-r)$}}][(1-x)(1-x-y)+r],$ \\

\begin{figure}[h]
\begin{minipage}[t]{0.45\textwidth}
\centering
\includegraphics[width=5cm , angle=0]{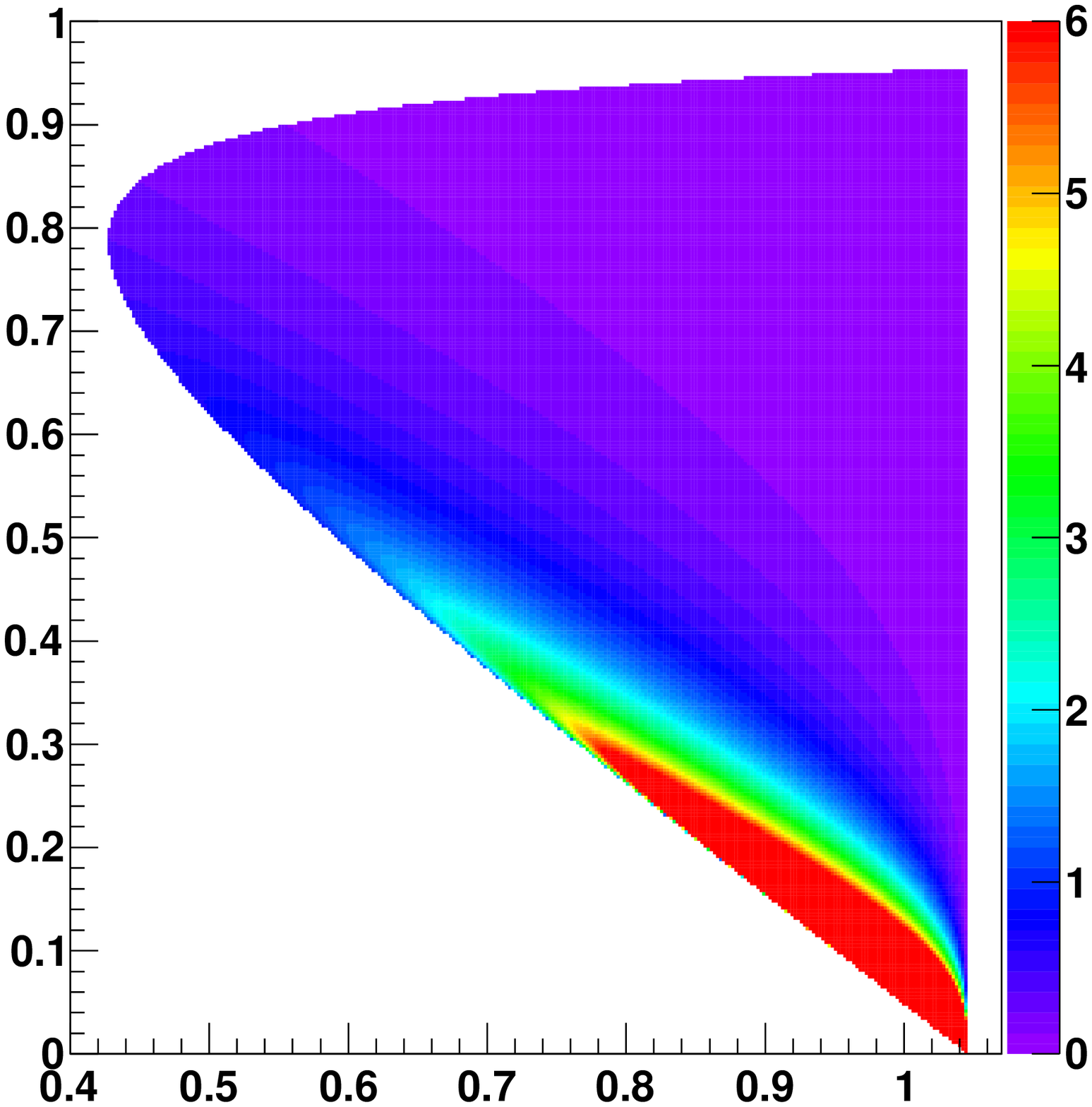}
\caption{IB.}\label{ib}
\end{minipage}  
\hspace{1cm}
\begin{minipage}[t]{0.45\textwidth}
\centering
\includegraphics[width=5cm , angle=0]{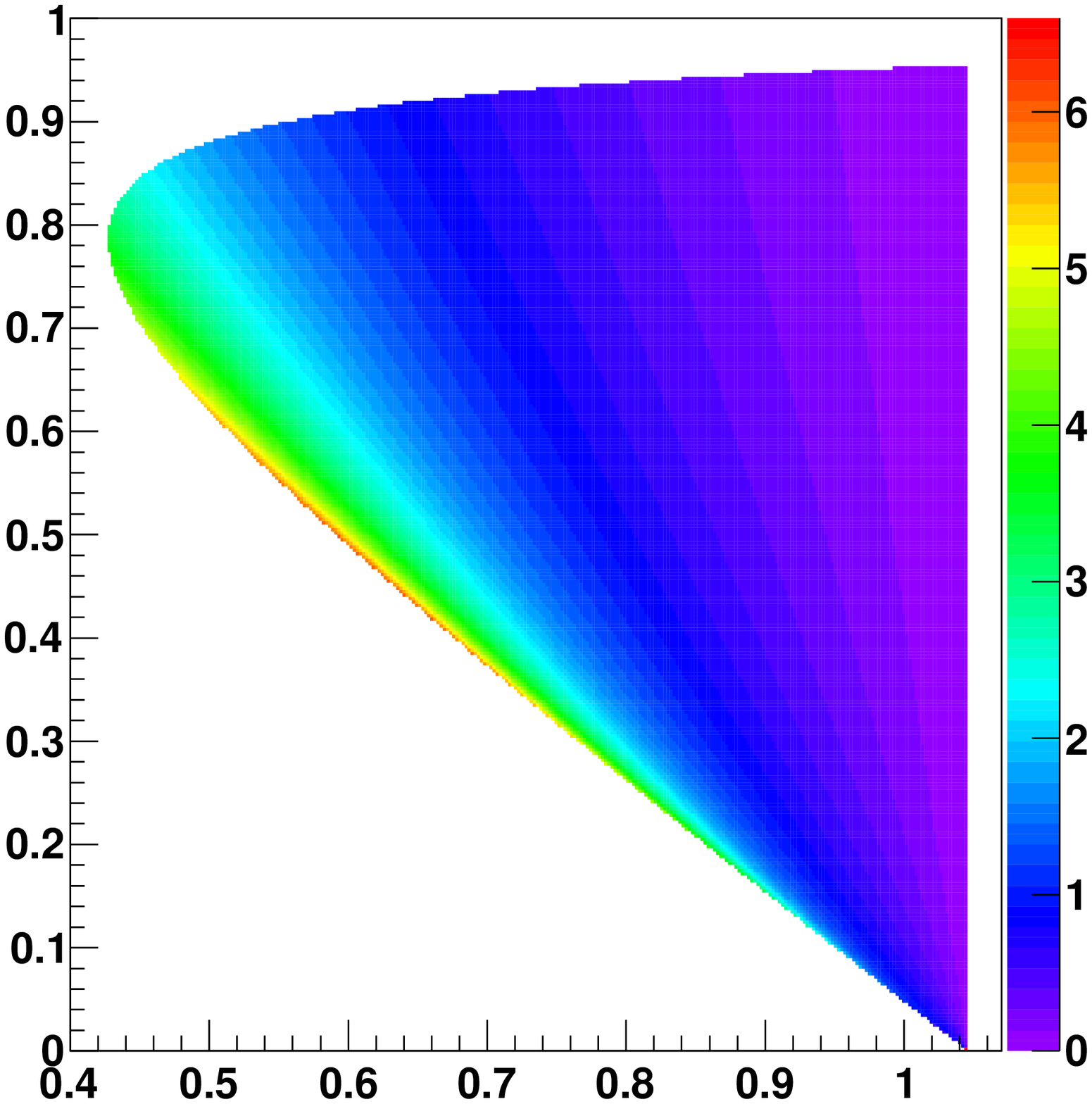}
\caption{INT$^-$.}\label{int-}
\end{minipage}
\end{figure}

\begin{figure}[h]
\begin{minipage}[t]{0.45\textwidth}
\centering
\includegraphics[width=5cm , angle=0]{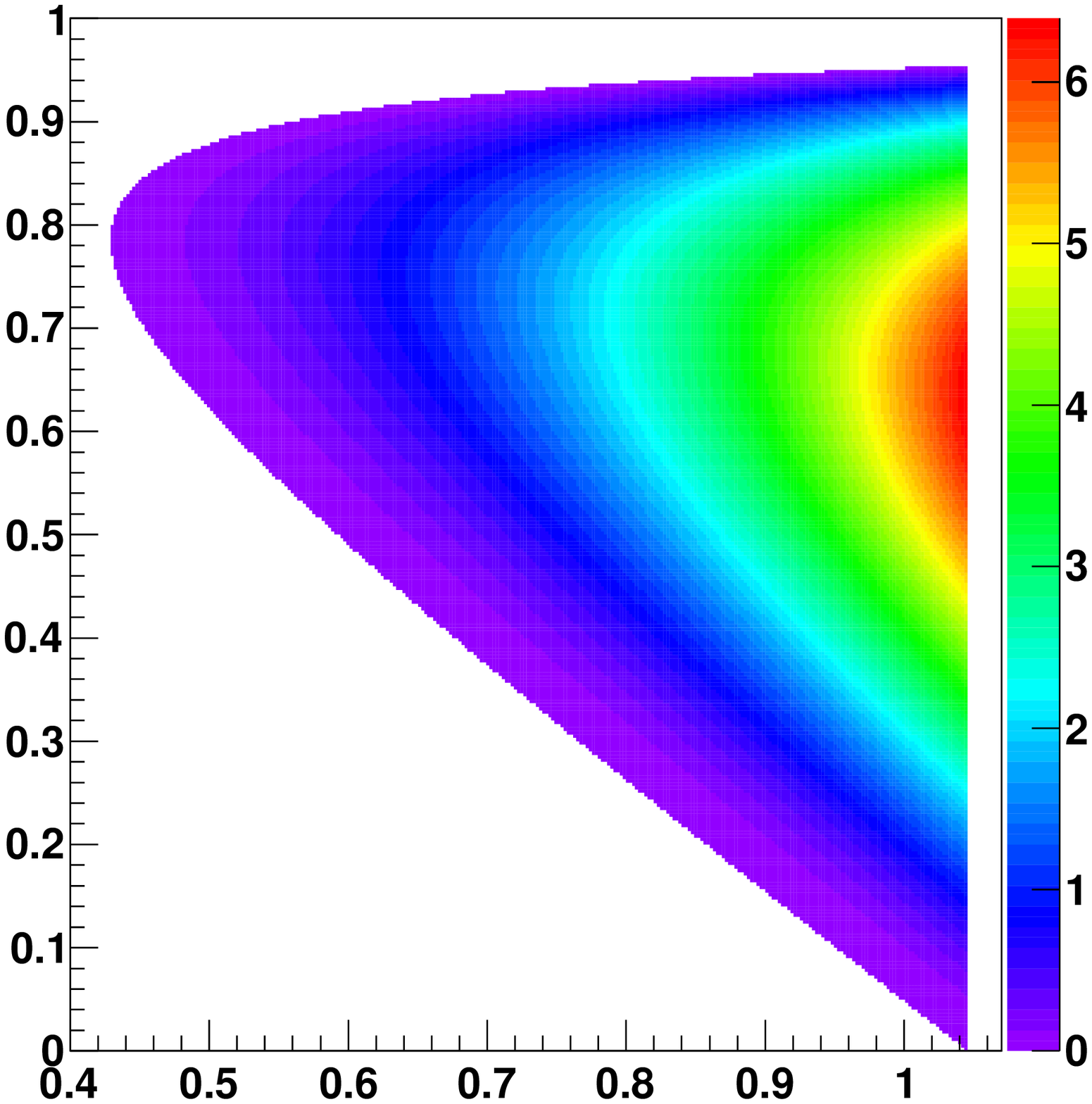}
\caption{SD$^+$.}\label{sd+}
\end{minipage}  
\hspace{1cm}
\begin{minipage}[t]{0.45\textwidth}
\centering
\includegraphics[width=5cm , angle=0]{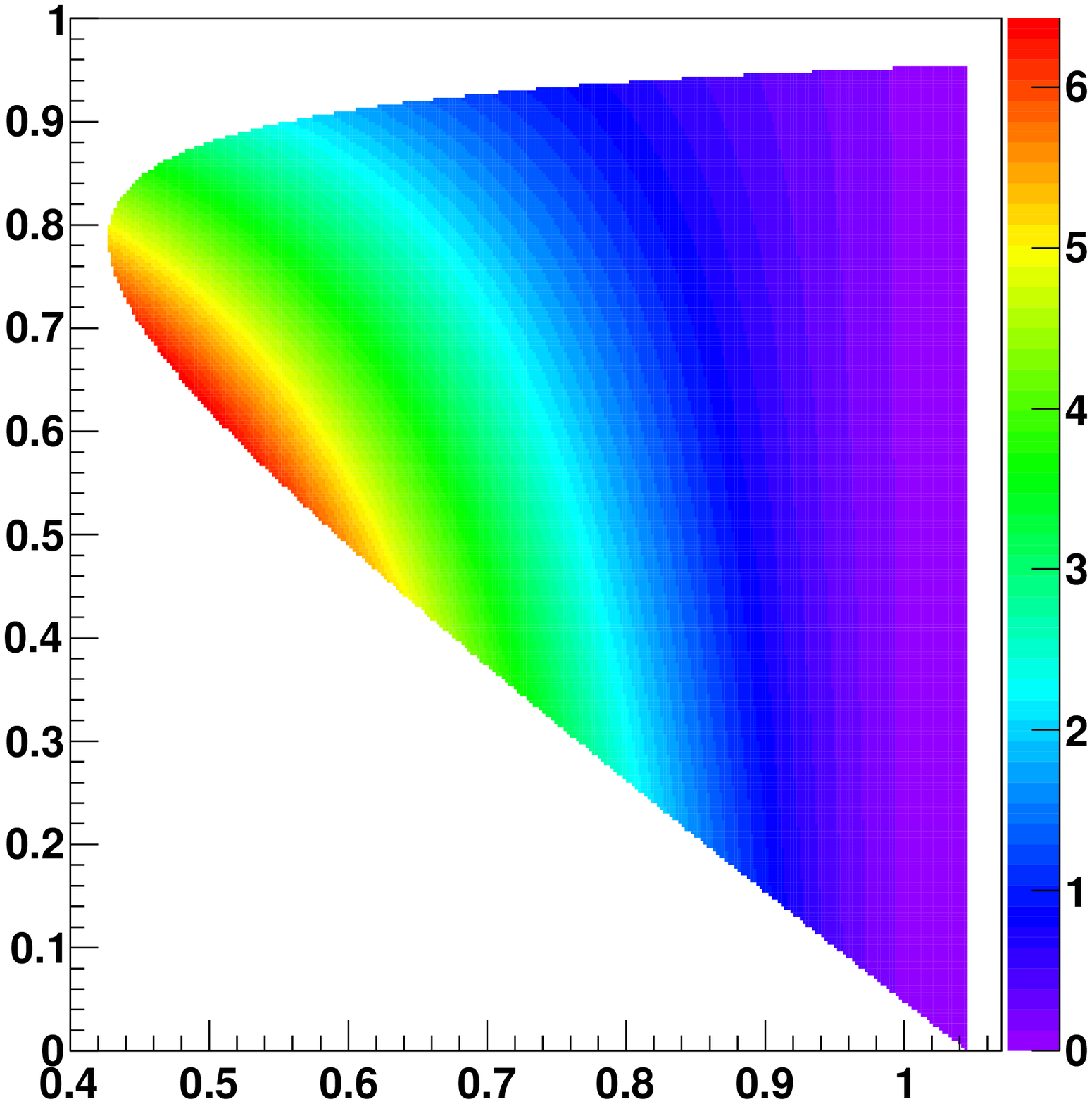}
\caption{SD$^-$.}\label{sd-}
\end{minipage} 
\begin{picture}(1,1)
\put(-50,0){$y$}
\put(-290,0){$y$}
\put(-50,205){$y$}
\put(-290,205){$y$}
\put(-185,130){$x$}
\put(-185,340){$x$}
\put(-425,130){$x$}
\put(-425,340){$x$}
\end{picture} 
\end{figure}

$f_{INT^-}(x,y)=[\frac{\mbox{$1-y+r$}}{\mbox{$x(x+y-1-r)$}}][x^2-(1-x)(1-x-y)-r], $\\

and 
$r= \left[\frac{\mbox{$M_{\mu}$}}{\mbox{$M_K$}}\right]^2$, 
$~A_{IB} = \Gamma_{K_{\mu2}}\frac{\mbox{$\alpha$}}{\mbox{$2\pi$}}\frac{\mbox{$1$}}{\mbox{$(1-r)^2$}}$,
$~A_{SD} = \Gamma_{K_{\mu2}}\frac{\mbox{$\alpha$}}{\mbox{$8\pi$}}\frac{\mbox{$1$}}{\mbox{$r(1-r)^2$}}\left[\frac{\mbox{$M_K$}}{\mbox{$F_K$}}\right]^2$,\\
$A_{INT} = \Gamma_{K_{\mu2}}\frac{\mbox{$\alpha$}}{\mbox{$2\pi$}}\frac{\mbox{$1$}}{\mbox{$(1-r)^2$}}\frac{\mbox{$M_K$}}{\mbox{$F_K$}}$, 
In these formulae, $\alpha$ is the fine structure constant, F$_K$ is $K^+$ decay constant $(F_K=156.1\pm0.2\pm0.8\pm0.2 MeV$ \cite{pdg})
 and $\Gamma_{K_{\mu2}}$ is $K_{\mu2}$ decay width.
Dalitz-plot distributions for different terms are shown in Figs.~\ref{ib}$-$\ref{sd-}.

$F_{V} \pm F_{A}$ are calculated within $\chi$PT
(O(p$^{4}$) \cite{rate}, O(p$^{6}$) \cite{chpt_p6}) and LFQM model \cite{lfqm}. In general, $F_{V}$ and $F_{A}$
depend on $q^2=(P_K-P_\gamma)^2=M_K^2(1-x)$.
In the O(p$^{4}$) $\chi$PT they are constant and $F_{V} + F_{A}=0.137$, $F_{V} - F_{A}=0.052$.
We will initially assume $F_{V}$ and $F_{A}$ constant and then test for their dependence on $q^2$.


 Experimentally, 
the decay $K\to \mu\nu_{\mu} \gamma$ was studied mostly in the IB dominated region (see \cite{mng2},\cite{mng3},\cite{mng4}).
There was only one formfactor measurement in E787
experiment \cite{mng1}. In this study, SD$^+$ term was extracted and $|F_{V}+F_{A}|$ was obtained to be $|F_{V}+F_{A}|=0.165\pm0.007(stat)\pm0.011(syst)$. 
Also $F_{V}-F_{A}$ was constrained: $-0.04 < F_{V}-F_{A} < 0.24$.
$F_{V}-F_{A}$ was measured by E865 experiment in $K \to \mu \nu e^{+} e^{-}$ decay \cite{e865}:
$F_{V}-F_{A}=0.077\pm0.028$. The goal of our study is to measure 
 $K^-\to\mu^-\bar\nu_\mu\gamma$ decay in the kinematic region where INT$^-$ term (and hence $F_{V}-F_{A}$) can be extracted.\\

\section{ISTRA+ experiment}
\subsection{Experimental setup}
\begin{center}
\begin{figure}[h]
\includegraphics[scale=.6 , angle=90]{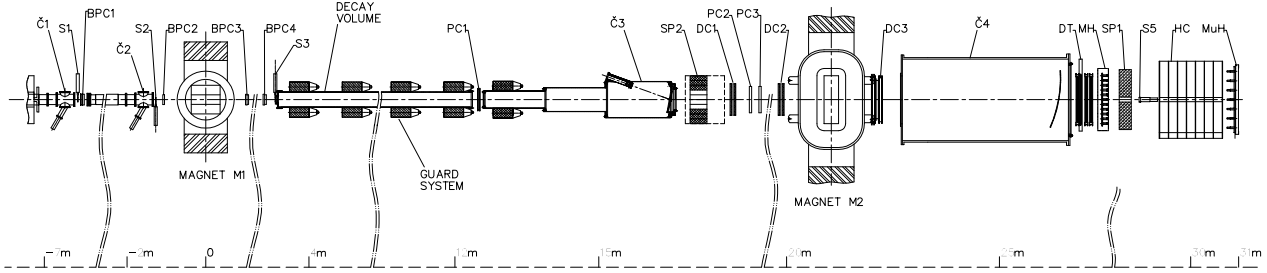}
\caption{ Elevation view of the ISTRA+ detector.}\label{detector}
\end{figure}
\end{center}
The experiment was performed at the IHEP 70 GeV proton synchrotron U-70.
The experimental setup ISTRA+ (Fig.~\ref{detector}) was described in details
in \cite{ISTRA}. 
 The setup was located in the  negative unseparated secondary beam. 
The beam momentum in the measurements was $\sim 26$ GeV/c with 
$\Delta p/p \sim 1.5 \% $. The admixture of $K^{-}$ in the beam was $\sim 3 \%$.
The beam intensity was $\sim 3 \cdot 10^{6}$ per 1.9 sec U-70 spill.
The track of a beam particle deflected by $M_{1}$ was 
measured by $BPC_{1}- BPC_{4}$ (1mm step multiwire chambers), the kaon identification was done by 
$\check{C_{0}} - \check{C_{2}}$ threshold $\check{C}$-counters. 
A 9 meter long vacuum
decay volume was surrounded by 8 lead glass rings $LG_{1} - LG_{8}$ used to
veto low energy photons. $SP_{2}$ was a lead glass calorimeter to detect/veto 
large angle photons.
Tracks of decay products deflected in $M_2$ with 1Tm field integral
were measured by $PC_{1} - PC_{3}$ (2mm step proportional chambers);
$DC_{1} - DC_{3}$ (1cm cell drift chambers) and finally by 2cm diameter
drift  tubes    $DT_{1} - DT_{4}$. 
Wide aperture threshold Cerenkov counters $\check{C_{3}}$, 
$\check{C_{4}}$ were filled  with He and
were not used in the measurements. Nevertheless $\check{C_{3}}$ was used as an extension of the decay volume.
$SP_{1}$ (ECAL) was a 576-cell lead glass calorimeter,
followed by HC (HCAL) - a scintillator-iron sampling hadron calorimeter. HC was subdivided
into 7 longitudinal sections 7$\times$7 cells each. MH was a 
11$\times$11 cell scintillating hodoscope used to  improve the time resolution of the 
tracking system, MuH  was a  7$\times$7 cell muon hodoscope. 

The trigger was provided by $S_{1} - S_{3}$, $S_5$ scintillation counters, 
$\check{C_{0}} - \check{C_{2}}$ Cerenkov counters,
analog sum of amplitudes from the last dinodes of the $SP_1$ :
 $T_{0}=S_{1} \cdot S_{2} \cdot S_{3} \cdot 
 \check{C_{0}} \cdot \bar{\check{C_{1}}} \cdot 
 \bar{\check{C_{2}}} \cdot 
 \bar{S_{5}} \cdot \Sigma(SP_{1})$,
here  $S_5$ was a counter downstream  the setup at the beam focus;
$\Sigma(SP_{1})$- a requirement for the analog sum of ECAL amplitudes 
to be above $\sim$3 GeV. The last requirement 
served to suppress the $K \rightarrow \mu \nu$ decay.
About $10\%$ events were recorded with a different trigger:
 $T_{1}=S_{1} \cdot S_{2} \cdot S_{3} \cdot 
 \check{C_{0}} \cdot \bar{\check{C_{1}}} \cdot 
 \bar{\check{C_{2}}} \cdot 
 \bar{S_{5}}$.
This prescaled trigger allowed to calculate trigger efficiency as a function of the energy released in ECAL.

\subsection{Data and MC samples}

We use high-statistics data collected in Winter 2001 run. About 332M events were stored on tapes.
This statistics was complemented by ~200M MC events generated with Geant3 \cite{geant3}. The MC generation included realistic description
of all ISTRA+ detectors.

\section{ Event selection}
\subsection{Selection criteria and general cuts}

The decay signature is defined as follows: one primary track (kaon),
 one negatively charged secondary track identified as muon;
 one shower in ECAL not associated with the charged track. Muon identification 
using ECAL and HCAL is described in our previous papers (\cite{kmu3-1},\cite{kmu3-2}).

Several cuts are applied to clean the data:
\begin{itemize}
\item number of beam and decay tracks in both $X$ and $Y$ projections is equal to 1;
\item CL (track quality) of primary tracks in both $X$ and $Y$ projections must be greater 
 than 10$^{-2}$;
\item CL of decay tracks is greater than 0.1 (decay-$X$) and 0.15 (decay-$Y$);
\item the angle between primary (kaon) and secondary (muon) track is greater than 2 mrad. 
\end{itemize}

The last cut eliminates most of undecayed beam particles.
The quality of decay track (described quantitatively by CL) is worse than that of beam track because of multiple scattering and detector resolution.

Cuts containing photon energy include:
\begin{itemize}
\item gamma energy in kaon rest frame is greater than 10 MeV;
\item no photons in SP$_2$ calorimeter (energy threshold is 0.5 GeV for total energy release);
\item no photons in GS.
\end{itemize}

For vertex characteristics we have the following requirements:
\begin{itemize}
\item z-coordinate must be within the interval 400 $<$ z$_{vtx}$ $<$ 1600cm;
\item (-3) $< x_{vtx} <$ 3cm;
\item (-2) $< y_{vtx} <$ 6cm;
\item CL of general vertex fit is greater than 10$^{-2}$.
\end{itemize}

Additional cuts are applied to suppress backgrounds:
\begin{itemize}
\item number of hits in matrix hodoscope (MH) is less than 3;
\item missing momentum $\overrightarrow p_{miss}=\overrightarrow p_K-\overrightarrow p_\mu-\overrightarrow p_\gamma$ 
does not point to the ECAL central hole (this cut effectively rejects background from $K\to\pi\pi^0$ decay when
 the lost photon from $\pi^0\to\gamma\gamma$ goes into the hole).
\end{itemize}

\subsection{Trigger efficiency}
\begin{figure}[h]
\centering
\includegraphics[width=6cm , angle=0]{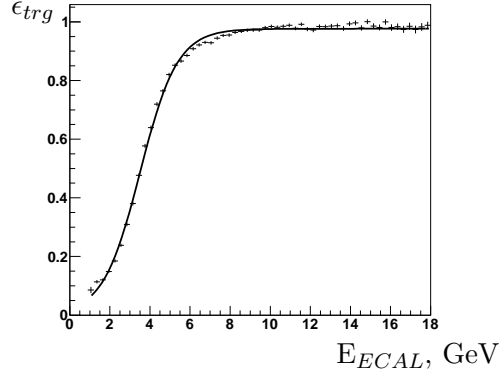}
\caption{$T_0$ trigger efficiency. Points - data, curve - fit by Fermi function.}\label{trig}
\begin{picture}(1,1)
\put(33,33){\small E$_{ECAL}$, GeV}
\put(-90,165){\small $\epsilon_{trg}$}
\end{picture}
\end{figure}

As $T_0$ trigger described in Section 2 contains energy threshold in SP$_1$ the trigger efficiency 
as a function of energy released in ECAL could be found using events with $T_1$ trigger:
$\epsilon_{trg}=(T_1\bigcap T_0)~/~ T_1$. Trigger curve is shown in the Fig.~\ref{trig}. The fit is done using Fermi function.
For the further analysis only events with $T_0$ are kept and these events are weighted by the factor of $1/\epsilon_{trg}$.

\section{Signal extraction}
\begin{figure}[h]
\begin{minipage}[t]{0.55\textwidth}
\centering
\includegraphics[width=6cm , angle=0]{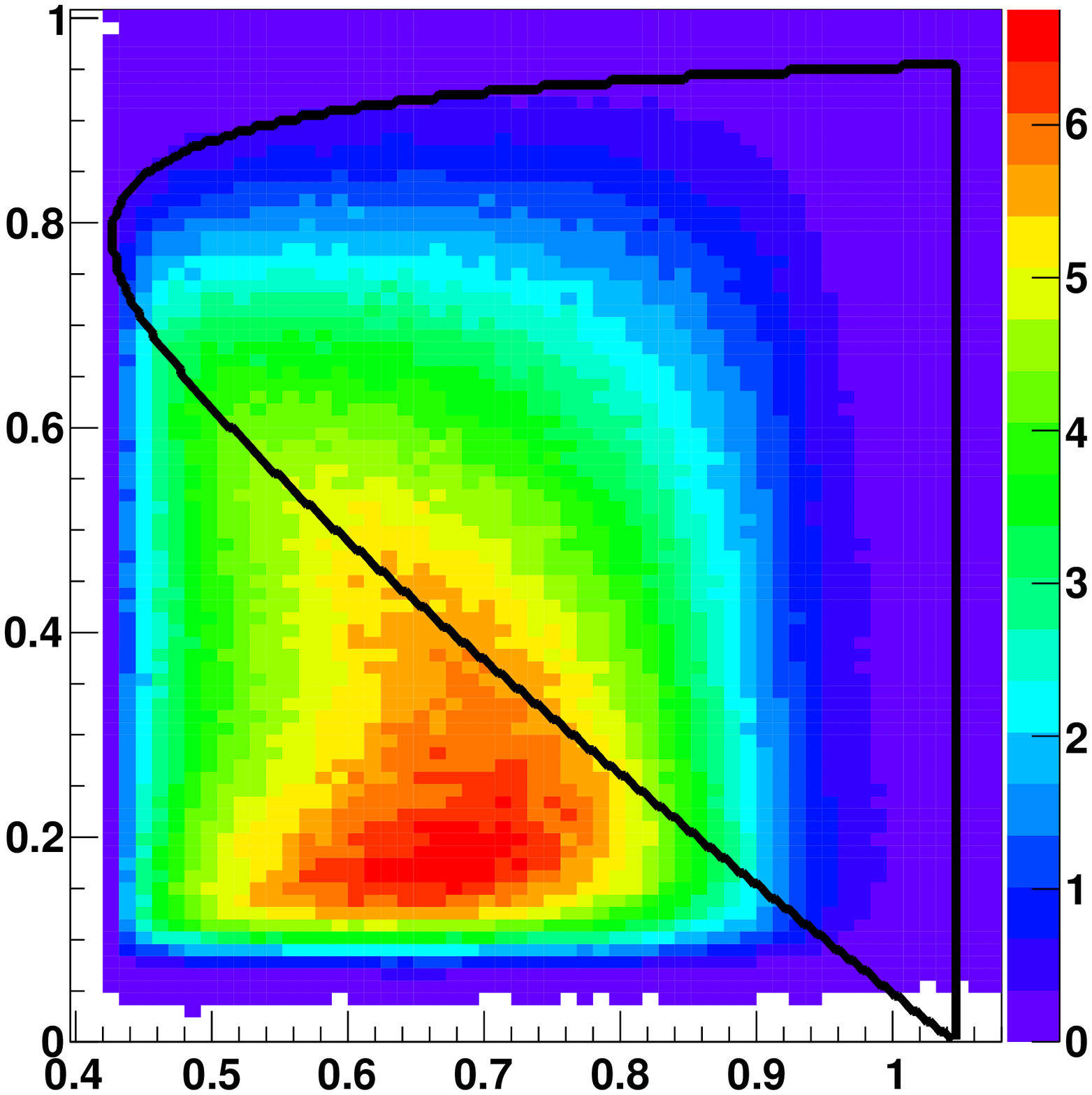}
\caption{Dalitz-plot density for $K\mu3$ bkg.}\label{bkgr1}
\end{minipage}  
\begin{minipage}[t]{0.45\textwidth}
\centering
\includegraphics[width=6cm , angle=0]{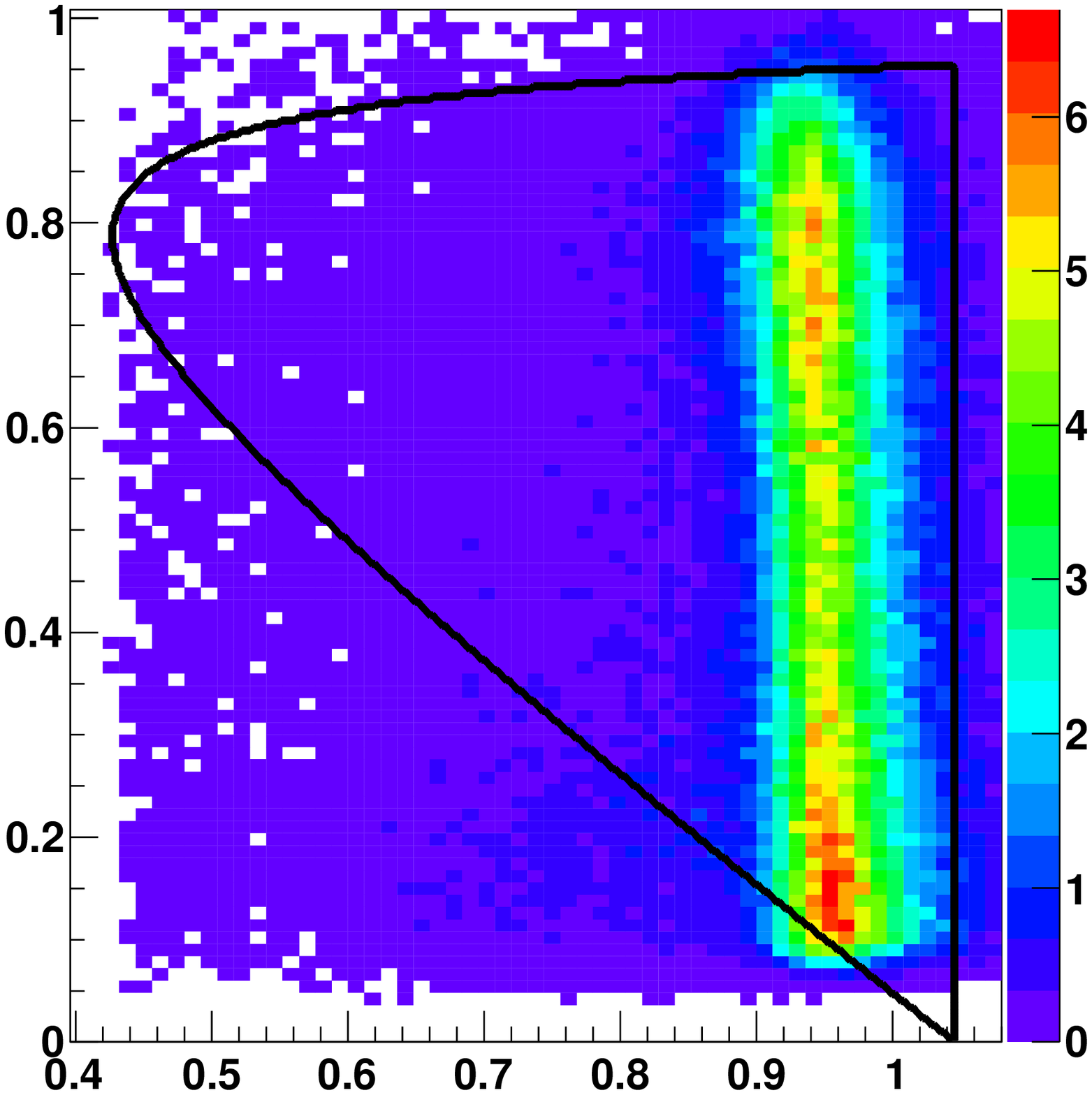}
\caption{Dalitz-plot density for $K\pi2$ bkg.}\label{bkgr2}
\end{minipage}
\begin{picture}(1,1)
\put(270,180){$x$}
\put(415,35){$y$}
\put(40,180){$x$}
\put(185,35){$y$}
\end{picture}
\end{figure}

 Distribution over $M(\mu\nu\gamma)$ is 
used for signal observation.
$M^2(\mu\nu\gamma)=(P_\mu+P_\nu+P_\gamma)^2$ where $P_\mu, P_\nu, P_\gamma$ are 4-momenta
of corresponding particles; missing mass m$_{\nu}$ is supposed to be equal to 0 so that
$\overrightarrow p_\nu=\overrightarrow p_K-\overrightarrow p_\mu-\overrightarrow p_\gamma; 
E_\nu=|\overrightarrow p_\nu|$. $M(\mu\nu\gamma)$ peaks at $K^-$ 
mass for the signal. Main background comes from 2 decay modes: $K^-\to\mu^-\nu\pi^0 (K\mu3)$
 and $K^-\to\pi^-\pi^0 (K\pi2)$ with one gamma lost from $\pi^0\to\gamma\gamma$
 and $\pi$ misidentified as $\mu$. Dalitz-plot distributions for $K\mu3$ and $K\pi2$ are shown in Figs.~\ref{bkgr1},~\ref{bkgr2}.

\subsection{Signal extraction procedure}

The procedure starts with dividing all kinematic ($x$,$y$) region into stripes on $x$ ($x$-stripes). The stripe width is $\Delta x$=0.05 ($\Delta E^*_{\gamma}\sim 24MeV$).
In every $x$-stripe we put a cut on $y$: 
$y_{min} < y < y_{max}$. $y_{min}$ and $y_{max}$ are selected from the maximization of signal significance defined as 
$\frac{S}{\sqrt{S+B}}$.

Besides distributions over $M(\mu\nu\gamma)$ and $y$, we use $cos~\theta^*_{\mu\gamma}$ for the signal observation, $\theta^*_{\mu\gamma}$ being the angle between $\vec p_{\mu}$
and   $\vec p_\gamma$ in c.m.s. We put a cut on $cos~\theta^*_{\mu\gamma}$ to reject background in those stripes where distributions over 
$cos~\theta^*_{\mu\gamma}$ for signal and background differ a lot (for example, in Fig.~\ref{fit1} in the
 the distribution over $cos~\theta^*_{\mu\gamma}$ signal peaks at 0.8 and background peaks at -0.4).  

\begin{figure}[h]
\begin{minipage}[t]{0.55\textwidth}
\centering
\includegraphics[width=6cm , angle=0]{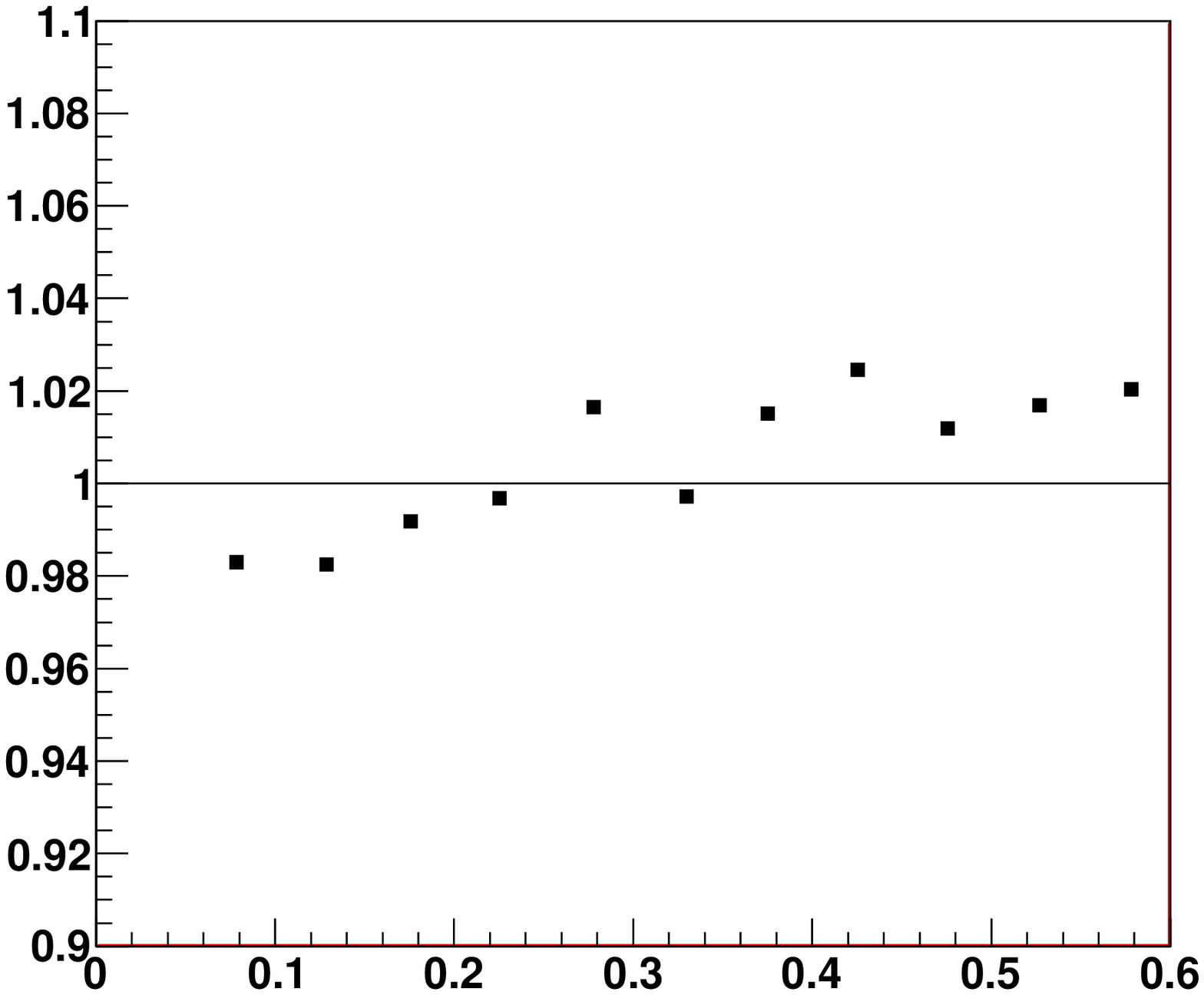}
\caption{Extra width $k_y$ vs $x$.}\label{sigma}
\end{minipage}  
\begin{minipage}[t]{0.45\textwidth}
\centering
\includegraphics[width=6cm , angle=0]{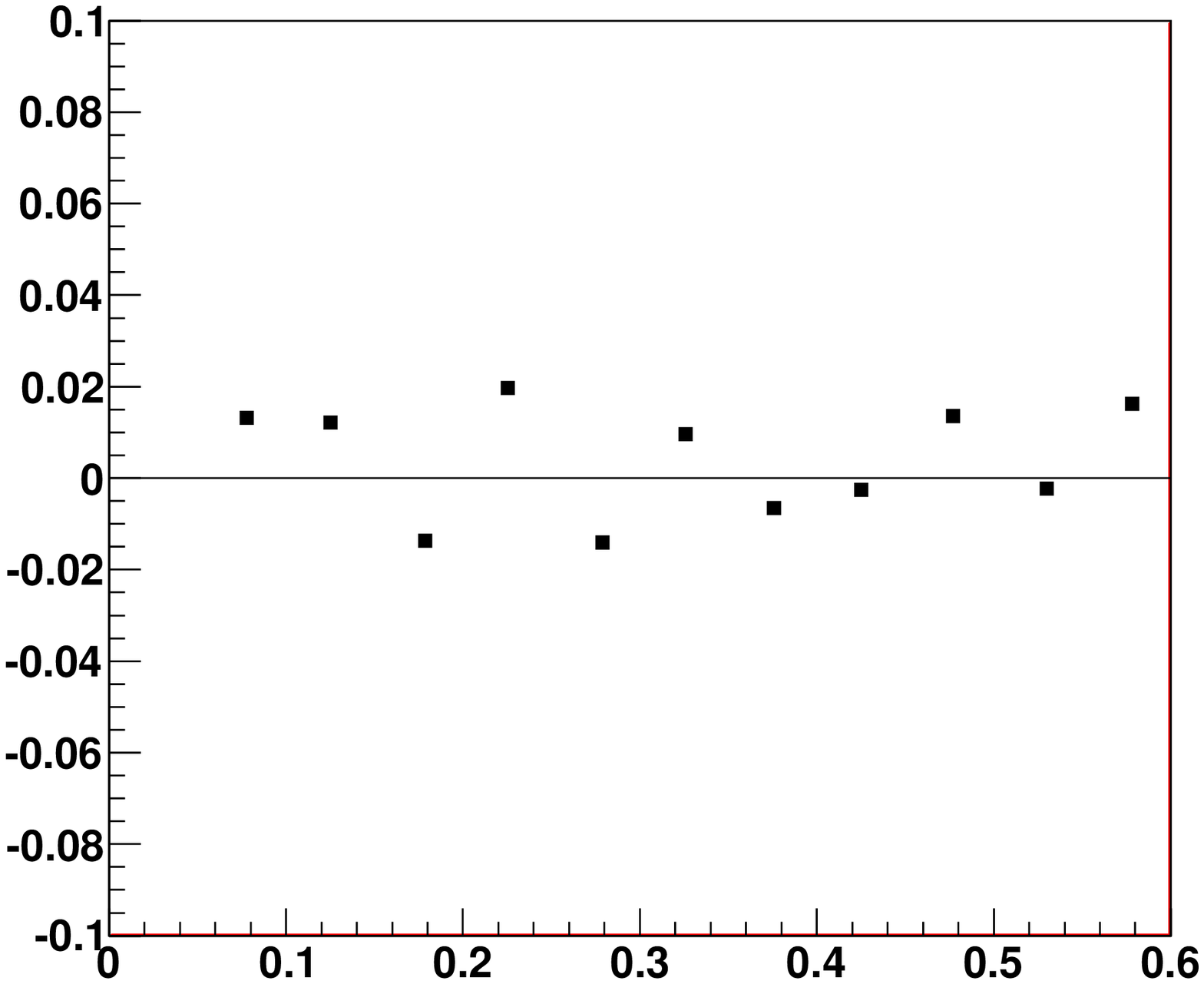}
\caption{Extra shift $b_y$ vs $x$.}\label{delta}
\end{minipage}
\begin{picture}(1,1)
\put(260,155){$b_y$}
\put(410,30){$x$}
\put(30,155){$k_y$}
\put(180,30){$x$}
\end{picture}
\end{figure}

Finally for each $x$-stripe we obtain events with cuts on $y$ and $cos~\theta^*_{\mu\gamma}$. Now we construct $M(\mu\nu\gamma$) which will be used for the fit.
 Fitting $M(\mu\nu\gamma$) alone is not sufficient because in some stripes distributions for signal and background are very similar. Also it would be difficult to 
distinguish between two backgrounds - $K\mu$3 and $K\pi$2.
That is why we take three histograms ($y$; $cos~\theta^*_{\mu\gamma}$ with cut on $y$; $M(\mu\nu\gamma)$  
with cuts on $y$ and $cos~\theta^*_{\mu\gamma}$) 
and fit them simultaneously. Both signal and background shapes are taken from MC. MC histograms are smoothed
and the result is stored as $f(z)$ function ($z=M(\mu\nu\gamma)$, $y$ or cos $\theta^{\star}_{\mu\gamma}$). 
For better fit we allow these functions to be slightly widen and shifted. We do it by using $f(k*z+b)$ instead of $f(z)$ in the fit, where fit parameters $k$ and $b$
are the same for signal and background and are different for $M(\mu\nu\gamma)$, $y$ and cos $\theta^{\star}_{\mu\gamma}$. In Figs.~\ref{sigma} and \ref{delta}
 extra width $k$ and extra shift $b$ for $y$-histogram are shown. For all selected $x$-stripes k$\sim$1 and b$\sim$0, i.e. our MC describes data properly.

The simultaneous fit gives signal event number in each $x$-stripe.
 As we use the same data several times we should take care about correct estimation of statistical error:
\begin{itemize}
\item do simultaneous fit of three histograms and obtain $\{p_i\}$ - best parameter values (they correspond to global $\chi^2$ minimum);
\item take $\{p_i\}$ as initial values and perform $\chi^2/n.d.f.$ and error estimation for one histogram $M(\mu\nu\gamma$) using MINOS program~\cite{minuit}.
\end{itemize}

\subsection{Selected kinematic region}

For further analysis we have selected eleven $x$-stripes in the following region: 
$0.05 < x < 0.6 ~(12MeV < E^\star_\gamma < 148MeV)$. The twelfth $x$-stripe is used for systematics study only.
Cuts on $y$ and $cos~\theta^*_{\mu\gamma}$ are summarized in table.\ref{cuts}.

\begin{table}[h]\footnotesize
\centering
    \begin{tabular}{|l|l|l|l|l|}  
      \hline  
   
        strip  & cut on $x$ & cut on $y$  &  $\Delta y$  &  cut on $cos~\theta^*_{\mu\gamma}$\\
         \hline  
          
01& $0.05<x<0.1$ & 0.9$-$1.1    & 0.2  &  $>-0.8$ \\ 
02& $0.1<x<0.15$ & 0.9$-$1.1    & 0.2  &  $>-0.8$ \\ 
03& $0.15<x<0.2$ & 0.85$-$1.    & 0.15 &  $>-0.8$ \\ 
04& $0.2<x<0.25$ & 0.8$-$0.95   & 0.15 &  $>-0.2$ \\ 
05& $0.25<x<0.3$ & 0.75$-$0.9   & 0.15 &  $>-0.3$ \\ 
06& $0.3<x<0.35$ & 0.72$-$0.87  & 0.15 &  $>-0.4$ \\ 
07& $0.35<x<0.4$ & 0.65$-$0.85  & 0.2  &  $>-0.3$ \\ 
08& $0.4<x<0.45$ & 0.62$-$0.85  & 0.23 &  $>-0.5$ \\ 
09& $0.45<x<0.5$ & 0.57$-$0.8   & 0.23 &  $>-0.7$ \\ 
10& $0.5<x<0.55$ & 0.52$-$0.75  & 0.23 &  $\--$ \\ 
11& $0.55<x<0.6$ & 0.48$-$0.7   & 0.22 &  $ \--$ \\  
12& $0.6<x<0.65$ & 0.45$-$0.65   & 0.2 &  $ \--$ \\  
                    
	  \hline

       \end{tabular}  
	\caption{Cuts on $y$ and $cos~\theta^*_{\mu\gamma}$ in $x$-stripes.}\label{cuts}
\end{table}

\begin{figure}[h]
\begin{minipage}[t]{0.45\textwidth}
\centering

\includegraphics[trim=0mm 15mm 0mm 10mm, clip, width=6cm , angle=0]{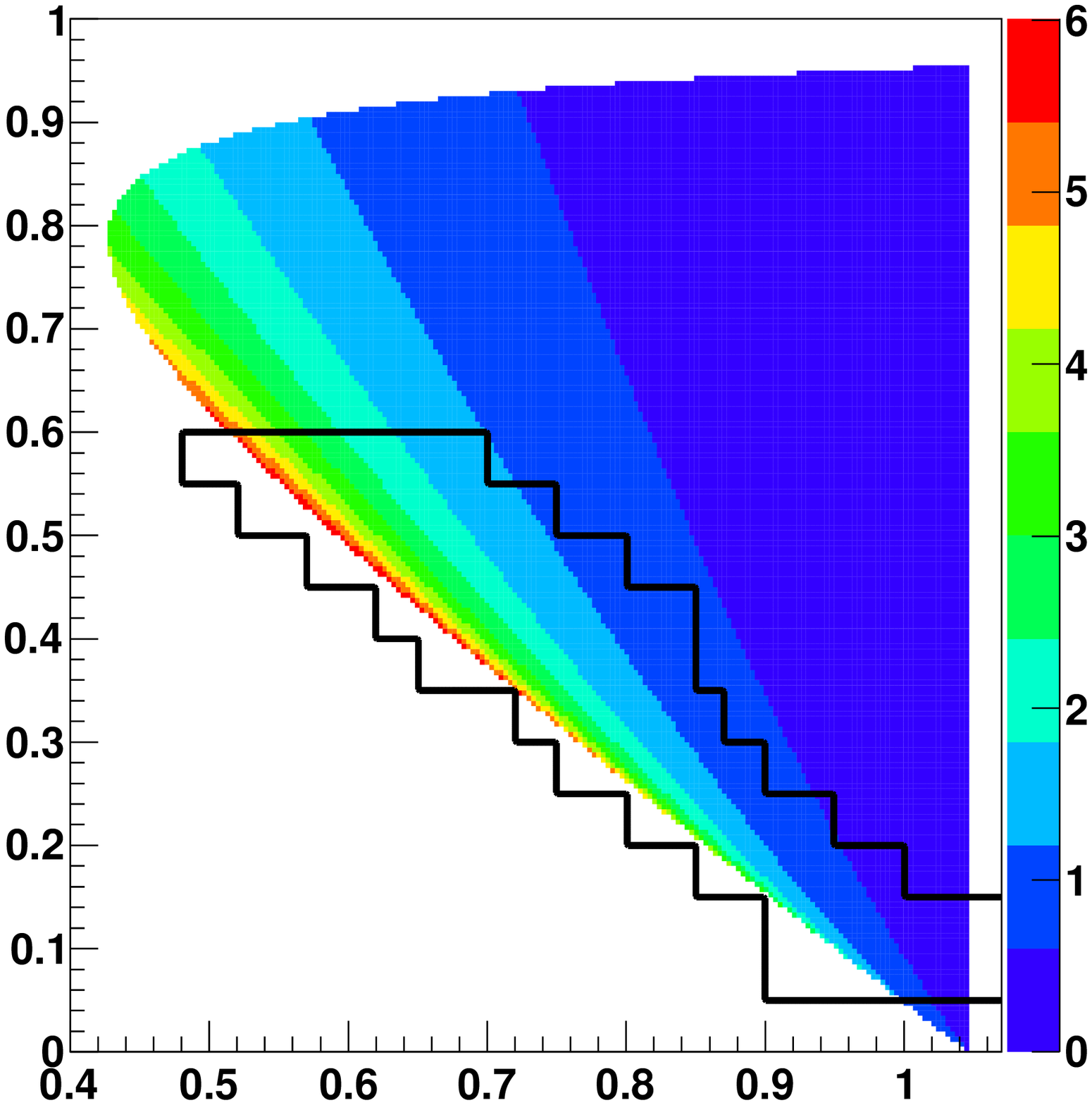}
\caption{INT$^-$ dalitz-plot density and selected kinematic region.}\label{mng}
\end{minipage}  
\hspace{1cm}
\begin{minipage}[t]{0.45\textwidth}
\centering

\includegraphics[width=5.5cm , angle=90]{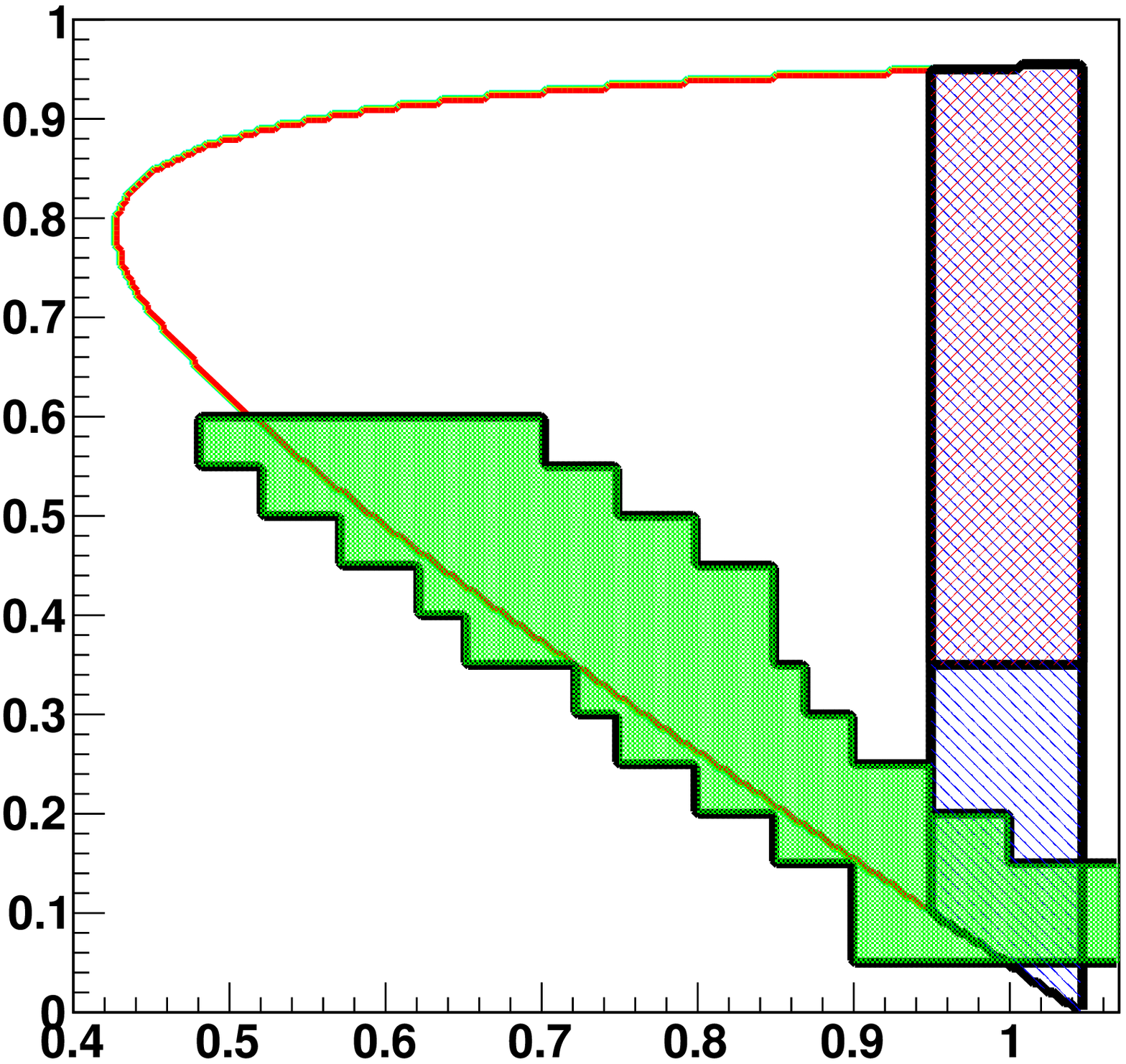}
\caption{ISTRA+(green); BNL E787(red hatch); KEK-104(blue hatch).}\label{mng2}
\end{minipage}
\begin{picture}(1,1)
\put(-435,147){$x$}
\put(-290,-10){$y$}
\put(-200,147){$x$}
\put(-40,-10){$y$}
\end{picture}
\end{figure}

   The $y$-width changes from stripe to stripe, in average $\Delta y\sim 0.2$.
 Our kinematic region is sensitive to INT$^-$ term (Fig.~\ref{mng}) and complementary to that of previous experiments \cite{mng1},\cite{mng2} 
(Fig.~\ref{mng2}). Stripe borders are slightly out of allowed kinematic region because of resolution.

\begin{figure}[h]
\begin{center}
\includegraphics[scale=.8 , angle=0]{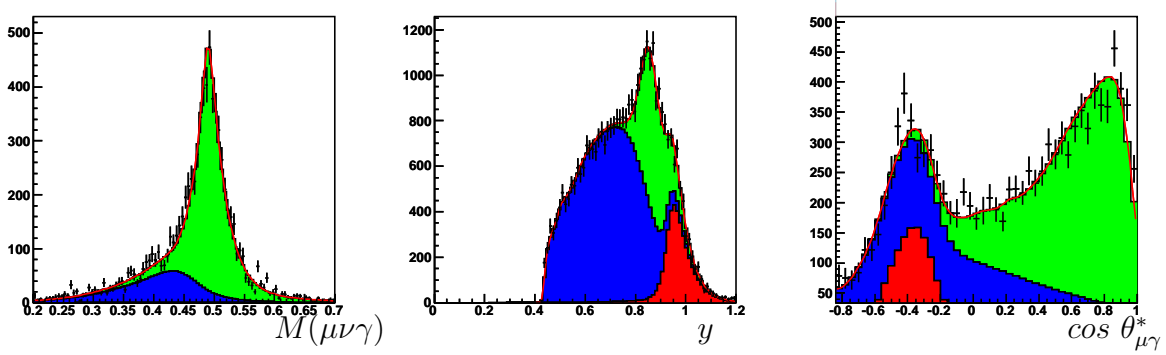}
\caption{Simultaneous fit in stripe 4: $M(\mu\nu\gamma)$, $y$ and $cos~\theta^*_{\mu\gamma}$.
Points with errors - data, blue - $K\mu3$, red - $K\pi2$, green - signal, red line - fit (signal+background). 
$\chi^2/n.d.f.$=157.3/91 (for mass histogram only, see text).}\label{fit1}
\end{center}
\begin{picture}(1,1)
\put(410,75){$cos~\theta_{\mu\gamma}^*$}
\put(270,75){$y$}
\put(110,75){$M(\mu\nu\gamma)$}
\end{picture}
\end{figure}

\begin{figure}[h]
\begin{center}
\includegraphics[scale=.8 , angle=0]{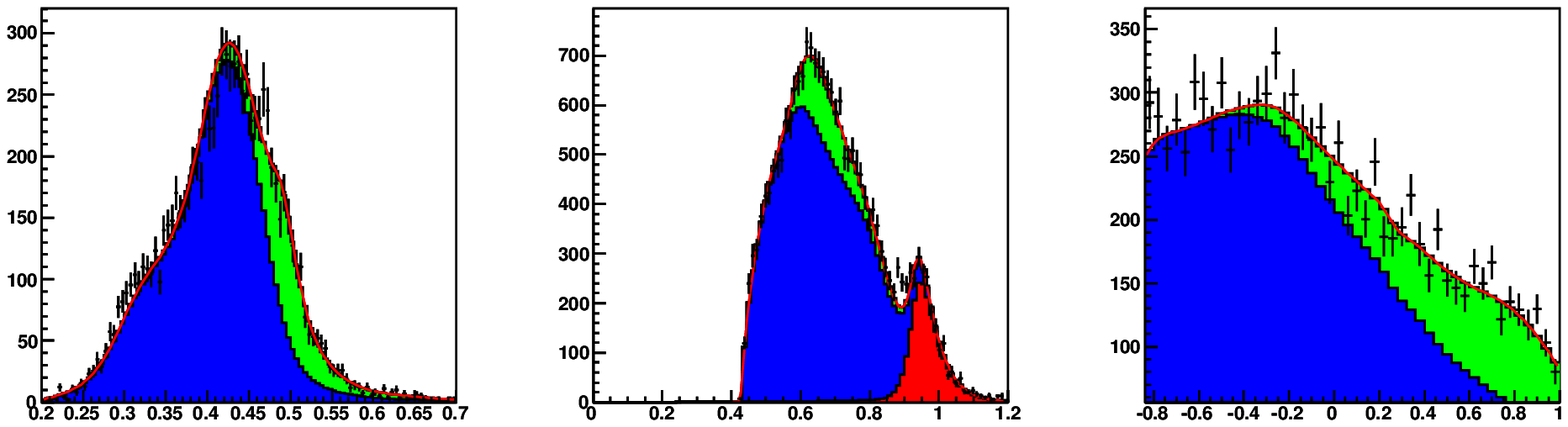}
\caption{Simultaneous fit in stripe 9: $M(\mu\nu\gamma)$, $y$ and $cos~\theta^*_{\mu\gamma}$.
Points with errors - data, blue - $K\mu3$, red - $K\pi2$, green - signal, red line - fit (signal+background). 
$\chi^2/n.d.f.$=141.7/91 (for mass histogram only, see text).}\label{fit2}
\end{center}
\begin{picture}(1,1)
\put(410,75){$cos~\theta_{\mu\gamma}^*$}
\put(270,75){$y$}
\put(110,75){$M(\mu\nu\gamma)$}
\end{picture}
\end{figure}

 Results of simultaneous fit for stripes $\# 4$ ($0.2 < x < 0.25$) and $\# 9$ ($0.45 < x < 0.5$) are shown in Figs.~\ref{fit1},~\ref{fit2}.
The total number of unweighted signal events is $\sim$22K.


\section{$F_{V}-F_{A}$ measurement}

\begin{figure}[h]
\begin{minipage}[t]{0.45\textwidth}
\centering
\includegraphics[width=6cm , angle=0]{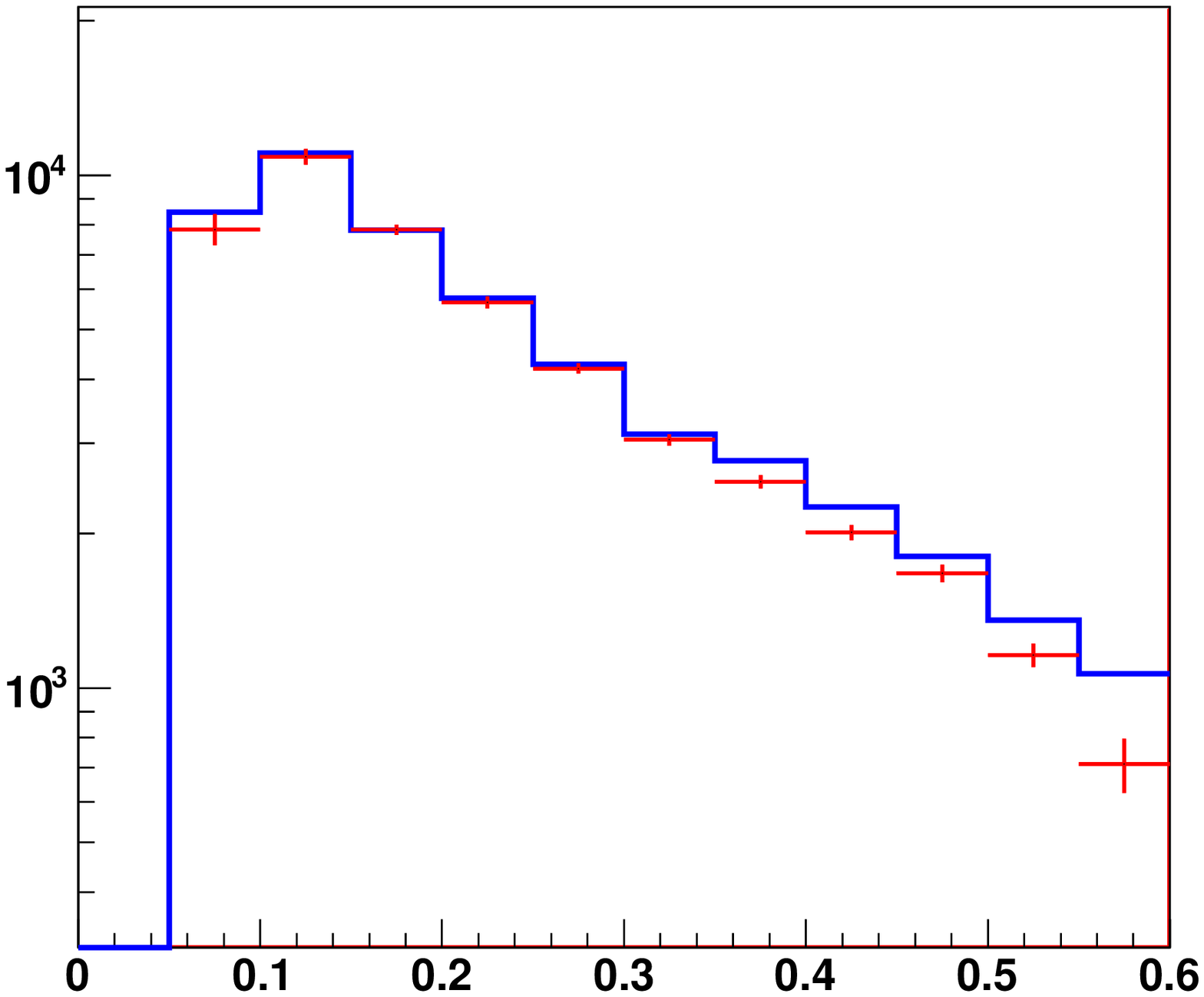}
\caption{$x$-spectrum. Red points - data, blue histogram - IB.}\label{x}
\end{minipage} 
\hspace{1cm}
\begin{minipage}[t]{0.45\textwidth}
\centering
\includegraphics[width=6cm , angle=0]{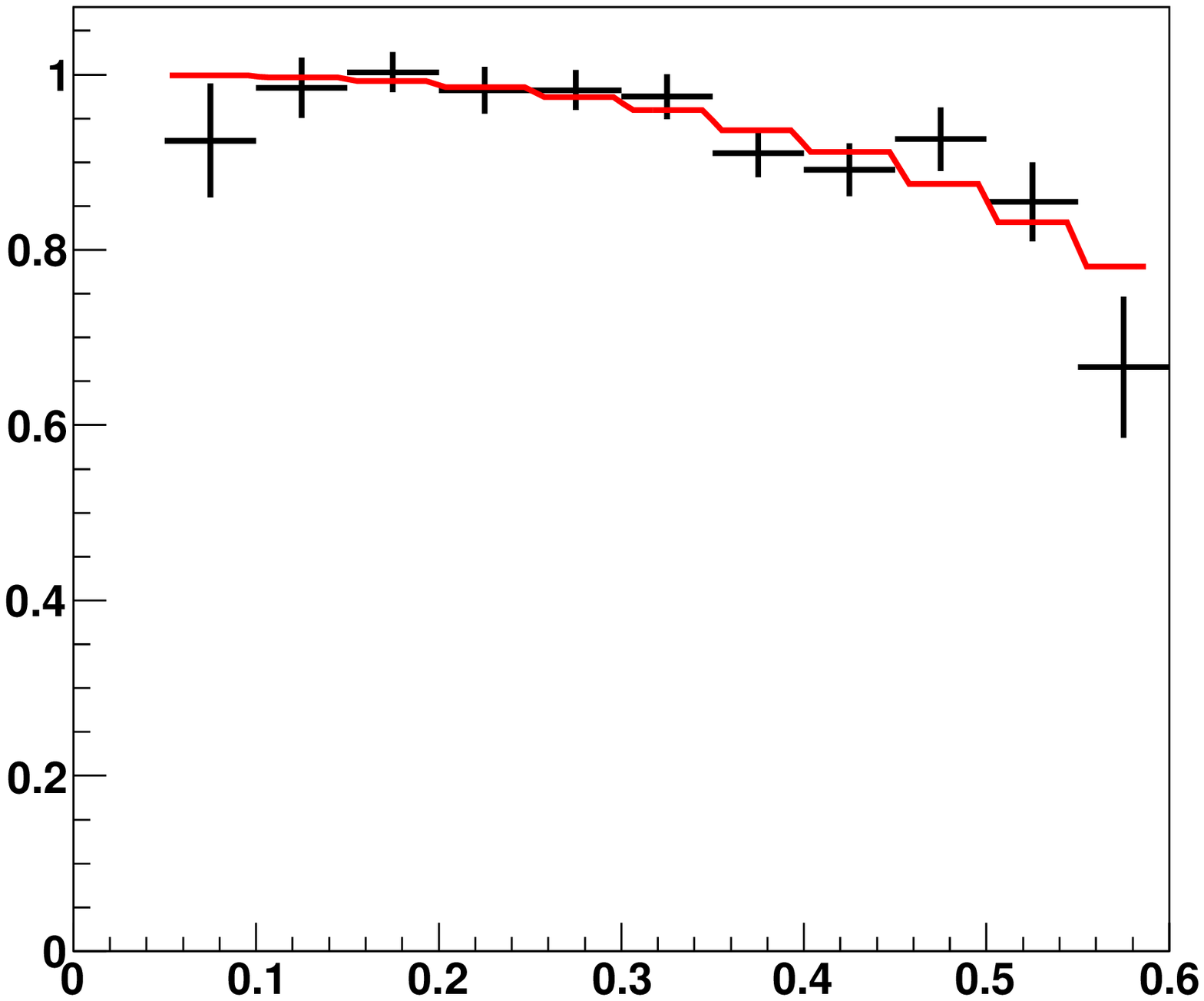}
\caption{N$_{exp}/N_{IB}$ for $x$-stripes and final fit. $\chi^2/n.d.f.$=7.8/9.}\label{ff}
\end{minipage}
\begin{picture}(1,1)
\put(-50,0){$x$}
\put(-290,0){$x$}
\put(-225,120){$N_{exp}/N_{IB}$}
\put(-445,120){$N_{exp}$}
\end{picture}
\end{figure}

For each $x$-stripe we have experimental event number $N_{exp}$ from fitting the data and IB event number $N_{IB}$
from MC (see Fig.~\ref{x}). Then we plot $N_{exp}/N_{IB}$ as a function of $x$ where each bin corresponds to a certain $x$-stripe (see Fig.~\ref{ff}).

For IB only we would have $N_{exp}/N_{IB}\approx 1$. It is the case for small $x$ where IB is dominated and INT$^-$ is negligible.
For large $x$ we see that $N_{exp}$ also contains negative interference term. We fit $N_{exp}/N_{IB}$ distribution with
$(f_{IB}(x)-f_{INT-}(x,p) )/f_{IB}(x)$ where $f_{IB}$ and $f_{INT-}$ give MC event number for a certain $x$-stripe and fit parameter p equals to $F_{V}-F_{A}$
($F_{V}$ and $F_{A}$ are initially assumed to be constant). 
The result of the final fit is as follows: $F_{V}-F_{A}=0.21\pm0.04(stat)$. Number of 'missing events' due to negative INT$^-$ term is 1483 
which is $\sim3\%$ of expected IB contribution (49722 weighted events). 

\section{Systematic error estimation}
The main potential sources of systematic error are:
\begin{itemize}
\item possible non-ideal description of signal/background shape in the simultaneous fit;
\item cut on $x$ (i.e. number of selected $x$-stripes);
\item $x$-binning (i.e. $x$-stripe width);
\item cut on $y$ in $x$-stripes;
\item cut on z-coordinate of the vertex;
\item possible contribution of INT$^+$ term.
\end{itemize}

Each source is investigated separately and errors are considered to be independent.

{\bf Possible non-ideal description of signal/background shape in the simultaneous fit}.
For estimation of shape systematics we scale errors in each bin of the final fit of Fig.~\ref{ff} 
proportionally to $\sqrt{\chi^2}$ ($\chi^2$ is obtained from simultaneous fit in a bin). Then we repeat final fit.
New value of $F_V-F_A$ is consistent with the main one and the fit error is larger:  $\sigma_{fit}\sim5\times10^{-2}$. 
We treat $\sigma_{fit}$ as follows: $\sigma_{fit}=\sqrt{\sigma_{stat}^2+\sigma_{syst,fit}^2}$ with $\sigma_{syst,fit}$ being systematical
error caused by non-ideal shape of signal and background distributions: $\sigma_{syst,fit}\sim3\times10^{-2}$. 

{\bf Cut on $\bf x$}. Each $x$-stripe has the width $\Delta x$=0.05. By adding/removing stripes involved in the fit on the left(right) border 
and repeating final fit we can see how $F_{V}-F_{A}$ depends on the $x$-cut value. 
For the left border, we take results of 3 fits which include stripes 1$-$11(main fit), 2$-$11 and 3$-$11. For the right border, we choose
fits including stripes 1$-$10, 1$-$11(main fit), 1$-$12. 

\begin{figure}[h]
\begin{minipage}[t]{0.45\textwidth}
\centering
\includegraphics[width=6cm , angle=0]{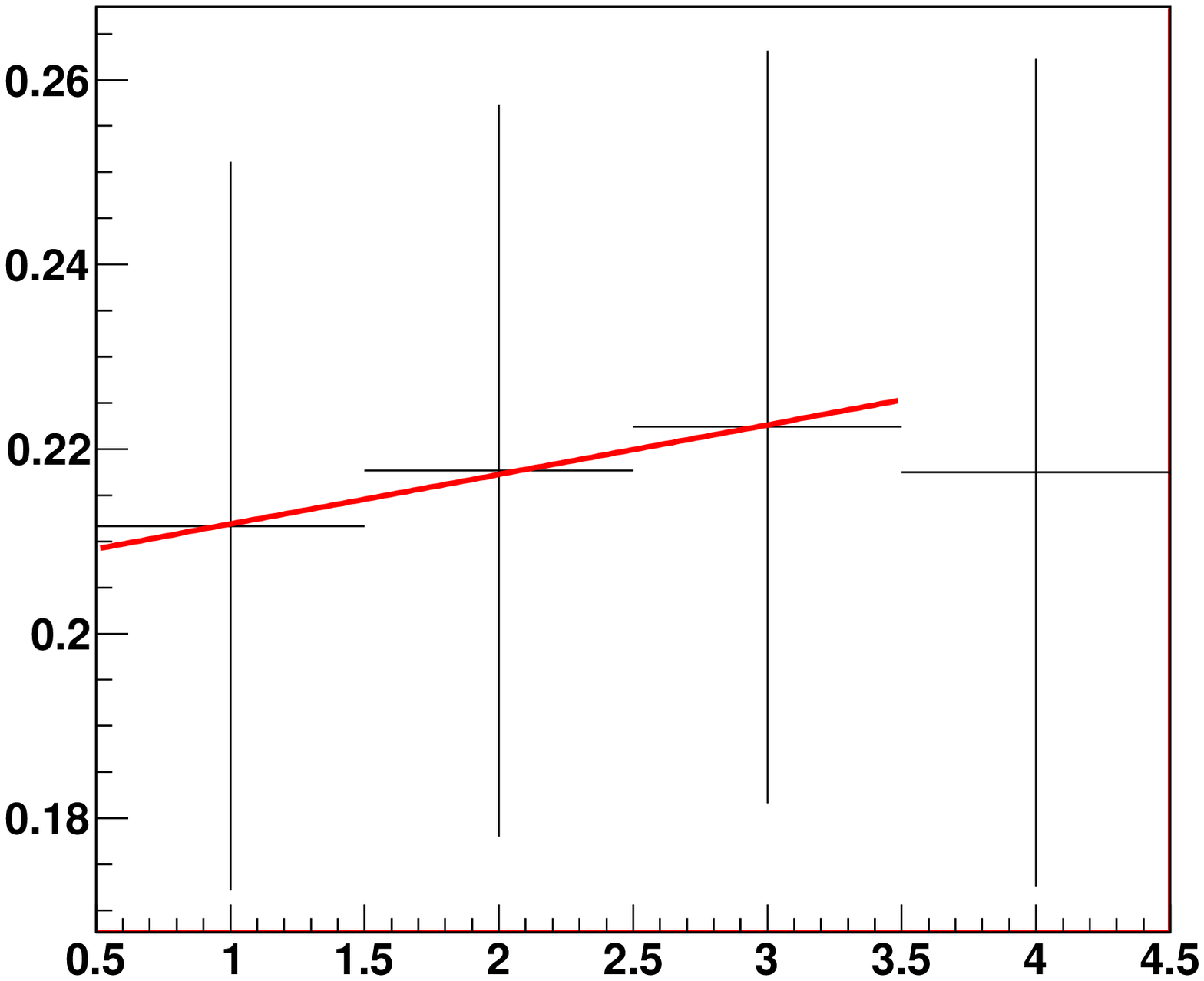}
\caption{Systematics of cut on $x$. Left border.}\label{left}
\end{minipage} 
\hspace{1cm}
\begin{minipage}[t]{0.45\textwidth}
\centering
\includegraphics[width=6cm , angle=0]{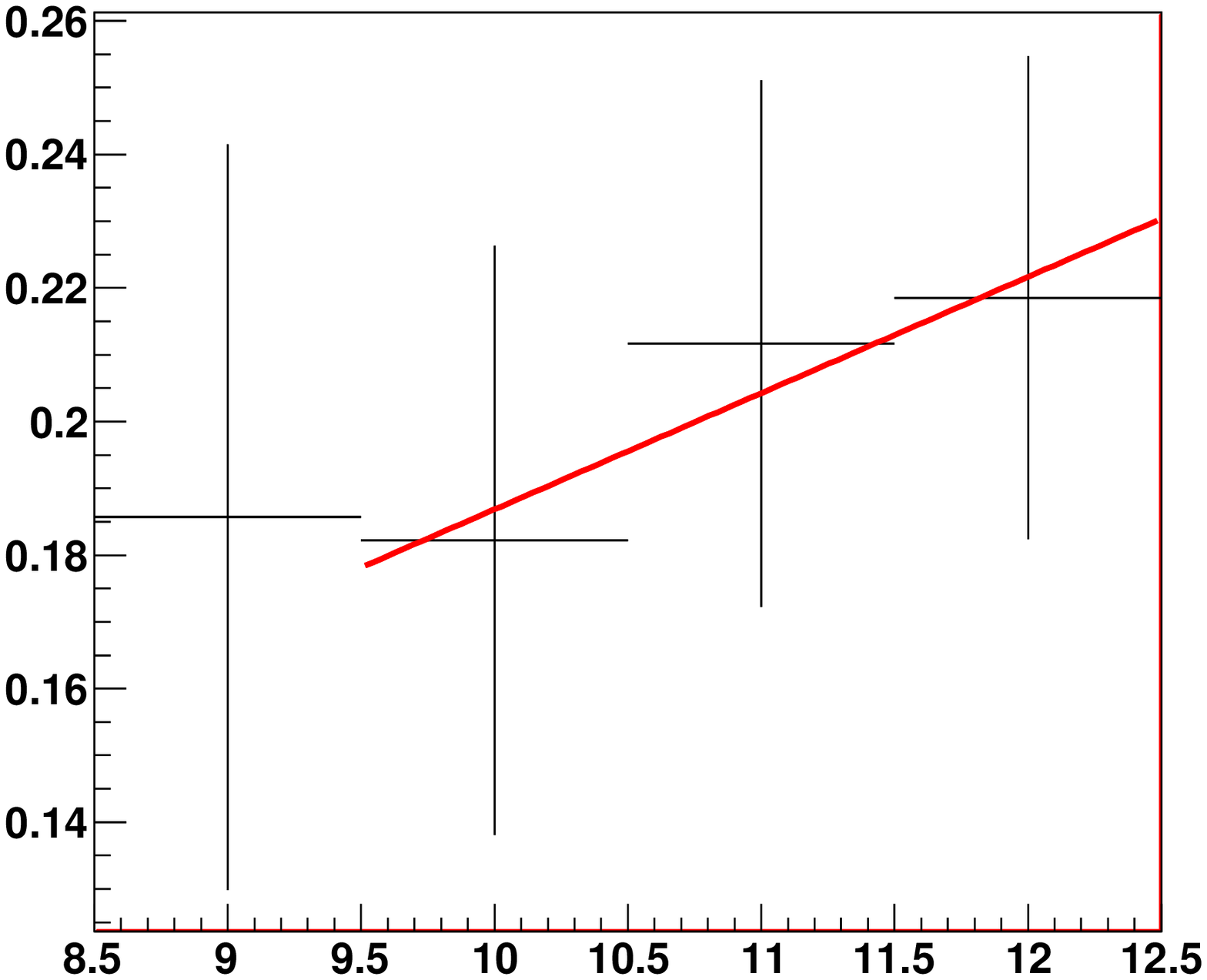}
\caption{Systematics of cut on $x$. Right border.}\label{right}
\end{minipage}
\begin{picture}(1,1)
\put(-70,0){$x$-stripe number}
\put(-310,0){$x$-stripe number}
\put(-230,125){$F_{V}-F_{A}$}
\put(-470,125){$F_{V}-F_{A}$}

\put(-395,30){\tiny main}
\put(-395,23){\tiny fit}
\put(-95,48){\tiny main}
\put(-92,41){\tiny fit}
\end{picture}
\end{figure}

The resulting plots $F_{V}-F_{A}$ vs $x$-cut are shown in Figs.~\ref{left} and \ref{right}.
 For the conservative estimate of systematics we fit these plots with straight
lines. The line slope multiplied by the resolution in $x$ (which is taken from MC) gives systematic error of this cut.
The systematic error of the right border is found to be $\sim 1.2\times 10^{-2}$ and that of the left border is negligible.

{\bf $\bf X$-binning}.
Changing width of $x$-stripes could shift the result of the final fit.
To test that, we repeat the whole procedure (choosing cuts on $y$ in $x$-stripes, simultaneous fits and a final fit)
for two different values of $\Delta x$: $\Delta x$=0.035 and $\Delta x$=0.07.
The smaller value of $\Delta x$ equals resolution in $x$. By comparing results obtained in the fits with new $\Delta x$
with the main one ($\Delta x$=0.05) we find $\epsilon_{syst} \sim 2\times 10^{-2}$.

{\bf Cut on $\bf y$ in $\bf x$-stripes}.
To investigate this source of systematics we choose cut on $y$ in a different way: instead of using significance we 
select events inside FWHM in $y$-distribution for signal MC.
 Such cuts on $y$ are stronger than those made using significance.
We redo simultaneous fit in $x$-stripes and final fit. The obtained result is consistent with the main one. No systematics is found here.

{\bf Cut on $\bf z$-coordinate of the vertex}.
To study this systematics we divide events into two groups - with $z_{vtx} < 1100$ cm and $z_{vtx} >$ 1100 cm.
The events with $z_{vtx} <$ 1100 cm use PC1 in the decay track reconstruction while events with $z_{vtx} >$ 1100 cm do not.
 Besides that the second group of events has the vertex 
inside the decay volume filled with He. It could be a possible source of systematics.
Repeating the whole procedure (simultaneous fit in $x$-stripes and final fit) we obtain two values for $F_{V}-F_{A}$ which are averaged.
The obtained values are compatible. No systematics is found here.

\begin{figure}[h]
\centering
\includegraphics[width=6cm , angle=0]{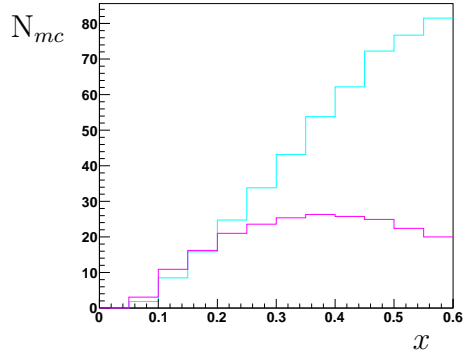}
\caption{INT$^+$(magenta) and INT$^-$(cian) contributions with $F_{V}+F_{A}$=0.165 and $F_{V}-F_{A}$=0.21.}\label{intp}
\begin{picture}(1,1)
\put(52,37){$x$}
\put(-100,155){N$_{mc}$}
\end{picture}
\end{figure}

{\bf Possible contribution of INT$^+$ term}. 
Contribution of INT$^+$ term is proportional to $F_{V}+F_{A}$. In E949 experiment\cite{mng1} only the absolute value of $F_{V}+F_{A}$ was measured. Fig.~\ref{intp}
shows that INT$^+$ could be rather large compared to INT$^-$, especially for small $x$. Fortunately, in this case INT$^-$ is negligible
with respect to IB. Since we know only  $|F_{V}+F_{A}|$
we add INT$^+$ term with both signs to the fitting function for the final fit and treat the difference in obtained value of $F_{V}-F_{A}$
as a systematic error. This way we get $\epsilon_{syst} \sim 1.4\times 10^{-2}$.

{\bf Total systematic error}.
Now we quadratically sum all sources supposing the errors to be independent and obtain
 $\epsilon_{syst} \sim 4\times 10^{-2}$. Contributions are summarized in table~\ref{syst}.

\begin{table}[h]\footnotesize
\centering
    \begin{tabular}{|l|l|}  
      \hline  
   
        {\bf source}  & {\bf systematic error}\\
         \hline  
          
non-ideal shape & 3$\times 10^{-2}$ \\ 
cut on $x$ & 1.2$\times 10^{-2}$ \\ 
$x$-binning & 2$\times 10^{-2}$ \\ 
cut on $y$ & -- \\ 
cut on $z_{vtx}$ & -- \\ 
INT$^+$ contribution & 1.4$\times 10^{-2}$ \\
 \hline
total &  4$\times 10^{-2}$ \\
                    
	  \hline

       \end{tabular}  
	\caption{Systematic errors.}\label{syst}
\end{table}

\section{Final result}

With this estimation of systematic error we finally get our  result:
$F_{V}-F_{A}=0.21\pm0.04(stat)\pm0.04(syst)$. 
It is $\sim3\sigma$ larger than theoretical prediction within $\chi$PT at O(p$^{4}$).

\begin{figure}[h]
\begin{minipage}[t]{0.45\textwidth}
\centering
\includegraphics[width=6cm , angle=0]{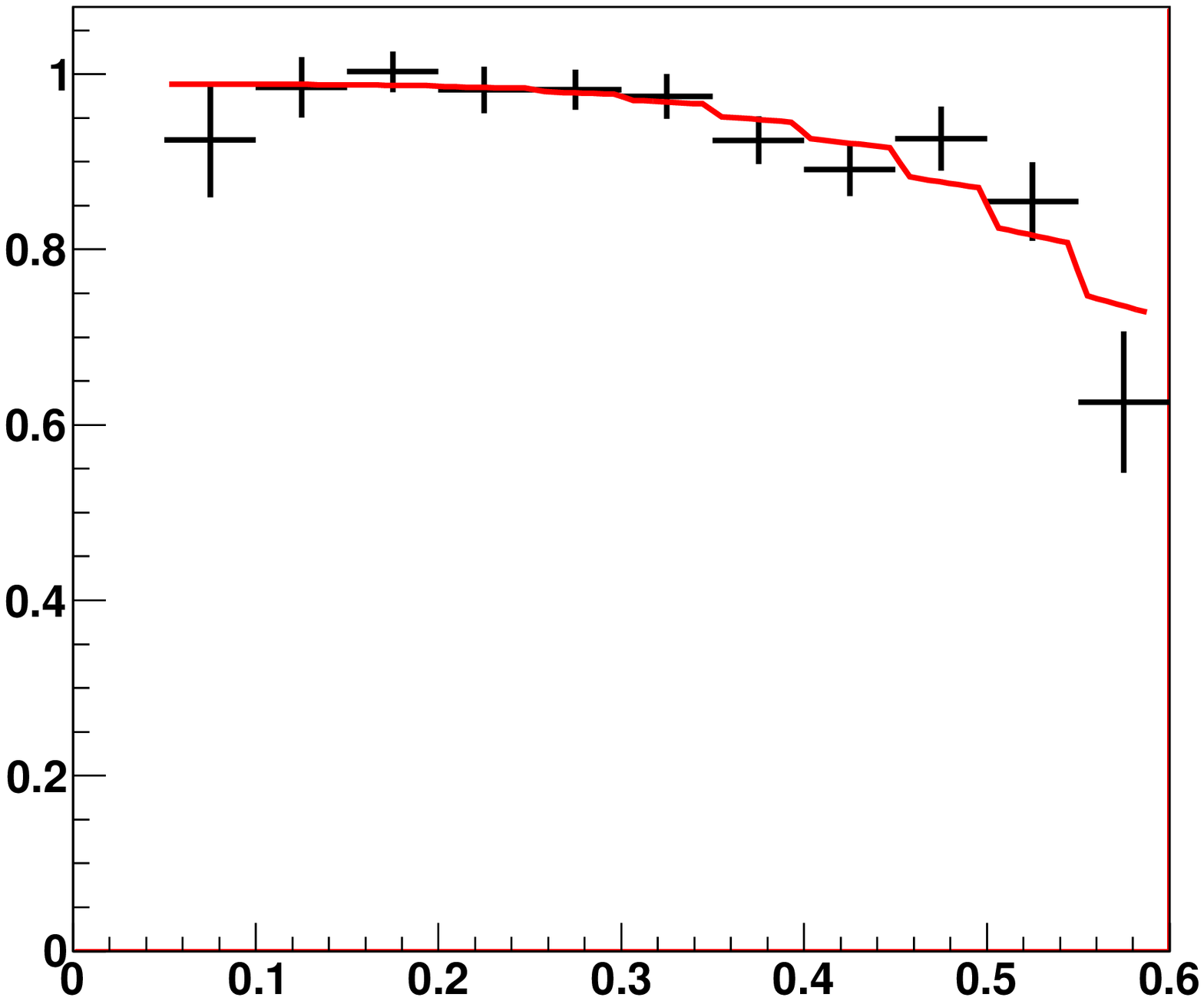}
\caption{$\chi PT ~O(p^6)$ fit, $F_V(0)$ and $F_A$ taken from theory. $\chi^2/n.d.f.$=8.8/9. }\label{fig_chptp6}
\end{minipage} 
\hspace{1cm}
\begin{minipage}[t]{0.45\textwidth}
\centering
\includegraphics[width=6cm , angle=0]{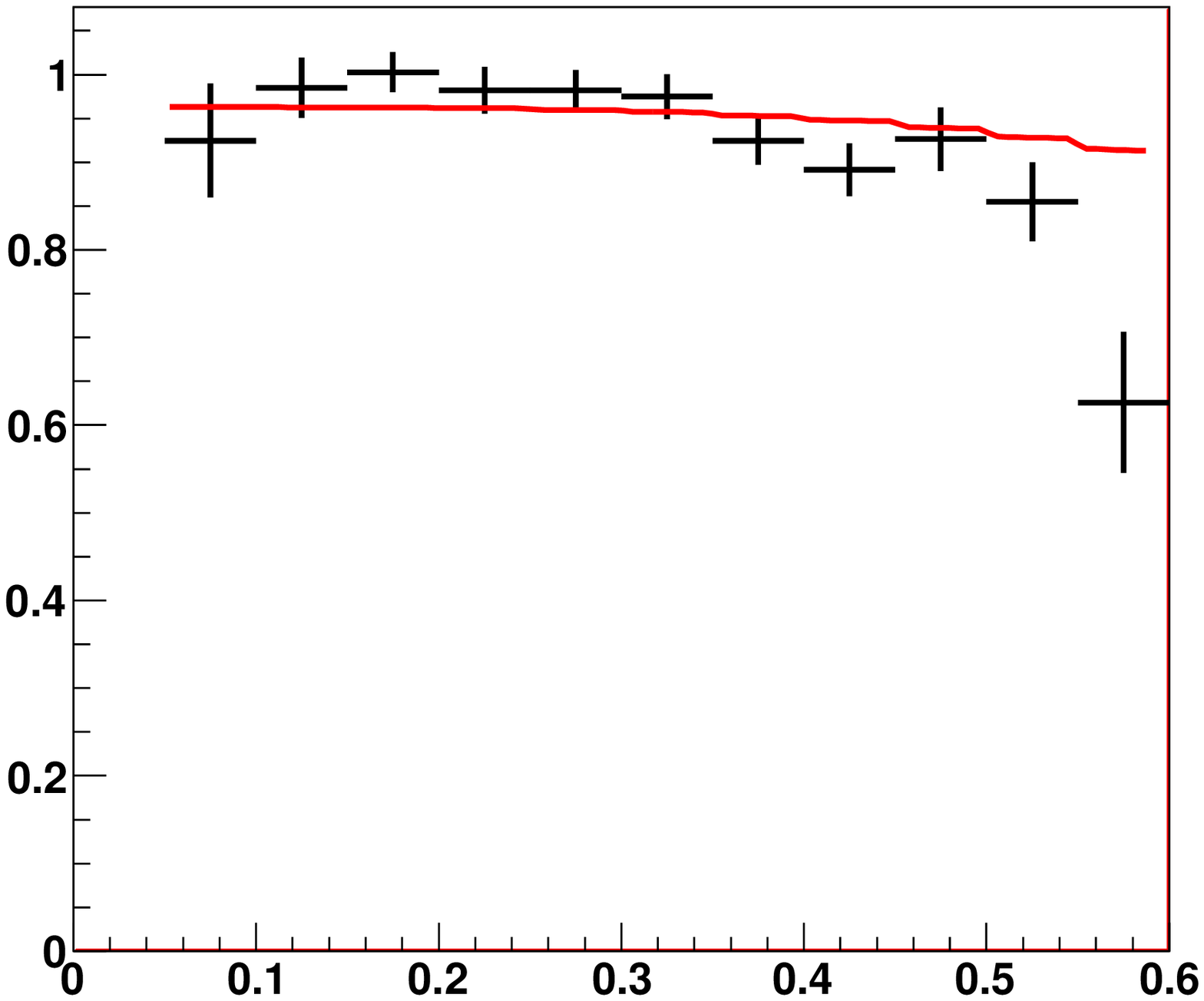}
\caption{LFQM fit, $F_V$ and $F_A$ taken from theory. $\chi^2/n.d.f.$=24.1/10.}\label{fig_lfqm}
\end{minipage}
\begin{picture}(1,1)
\put(-50,0){$x$}
\put(-290,0){$x$}
\put(-225,120){$N_{exp}/N_{IB}$}
\put(-465,120){$N_{exp}/N_{IB}$}

\end{picture}
\end{figure}

\begin{figure}[h]
\centering
\includegraphics[width=6cm , angle=0]{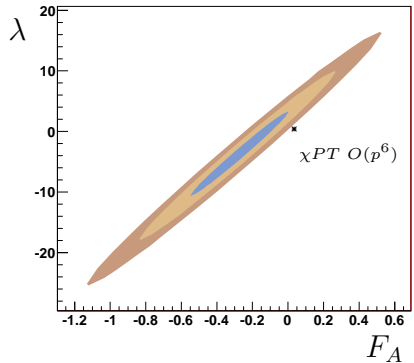}
\caption{$\chi PT ~O(p^6)$ fit, $F_V(0)$ taken from theory. $\chi^2/n.d.f.=7.5/8$.}\label{corr1}
\begin{picture}(1,1)
\put(50,35){$F_A$}
\put(-85,155){$\lambda$}
\put(25,110){\tiny $\chi PT ~O(p^6)$}
\end{picture}
\end{figure}

The $O(p^6)~ \chi PT $ gives linear dependence of $F_{V}$ and $F_{A}$ on $q^2$ (see \cite{chpt_p6}) and hence on $x$. 
We use $F_V$ and $F_{A}$ parametrization given in \cite{eng}:
$F_V=F_V(0)~[1+\lambda(1-x)]$, $F_A$=const.
This theoretical prediction was tested in three ways. First, we take both $F_{V}$ and $F_{A}$ from $O(p^6)~ \chi PT $ ($F_{V}(0)=0.082$,  $F_{A}$=0.034, $\lambda$=0.4)
 and do the final fit.
$\chi^2$ of this fit is 21.1/10 ($\sim 2.5 \sigma$ from $\chi^2=1$). Second, we take $F_V(0)$ and $F_A$ from $O(p^6)~ \chi PT$  
and take $\lambda$ as a fit parameter. It gives $\lambda=4.0\pm1.0$ with $\chi^2$=8.8/9 (Fig.~\ref{fig_chptp6}).
And finally we fix $F_V$(0) from $O(p^6) ~\chi PT$ and take $\lambda$ and $F_A$ as fit parameters. Correlation between them is shown in Fig.~\ref{corr1}.
Theoretical prediction is slightly out of 3$\sigma$-ellipse. 

In LFQM, $F_{V}$ and $F_{A}$ depend on $q^2$ in a complicated way (see \cite{lfqm}). Final fit is shown in Fig.~\ref{fig_lfqm}.
LFQM is disfavored ($\sim3 \sigma$ from $\chi^2=1$) although can not be excluded.

\section{Conclusions}
The radiative decay $K^-\to
\mu^-\bar\nu_{\mu} \gamma$   has been studied using in-flight decays at ISTRA+ setup.
 About 22K 
events  of $K^-\to\mu^-\bar\nu_{\mu} \gamma$ (it is the largest statistics for this decay) have been found in a 
new kinematic region. The negative INT$^-$ term has been observed and as a result $F_{V}-F_{A}$ has been measured:
 $F_{V}-F_{A}=0.21\pm0.04(stat)\pm0.04(syst)$. The result is $\sim 3 \sigma$ above $O(p^4)~ \chi$PT prediction.

An alternative analysis has been done by our collaboration (preliminary results are presented in \cite{istra_mng}). 
The results are in reasonable agreement ($\chi^2$=1.12).

Authors would like to thank C.Q.~Geng and E.~Goudzovsky for the code plotting formfactors in LFQM.
The work is supported by Russian Foundation for Basic Research (grants 10-02-00330 and 08-02-91016).

\end{document}